# How Can Generative AI Enhance the Well-being of Blind?

**Oliver Bendel**

School of Business FHNW
oliver.bendel@fhnw.ch

**Abstract**

This paper examines the question of how generative AI can improve the well-being of blind or visually impaired people. It refers to a current example, the Be My Eyes app, in which the Be My AI feature was integrated in 2023, which is based on GPT-4 from OpenAI. The author's tests are described and evaluated. There is also an ethical and social discussion. The power of the tool, which can analyze still images in an amazing way, is demonstrated. Those affected gain a new independence and a new perception of their environment. At the same time, they are dependent on the world view and morality of the provider or developer, who prescribe or deny them certain descriptions. An outlook makes it clear that the analysis of moving images will mean a further leap forward. It is fair to say that generative AI can fundamentally improve the well-being of blind and visually impaired people and will change it in various ways.

## Introduction

Some people are blind from birth. Others go blind in the course of their lives or develop more or less significant visual impairments and disabilities. Some train other senses to compensate and develop a particularly good sense of hearing or smell or use techniques such as echolocation for orientation (Yong 2023). Basically, the sense of sight is essential for humans to recognize their environment, to ward off dangers, and to perceive offers. Evolution has turned us into eye animals, and civilization and culture are mainly geared towards the sighted. With the invention of writing, but also of visual media such as photography and film, the latter development has continued dramatically. In such a world, one could say that the well-being of the blind and visually impaired is at risk.

The relevant WHO definition is as follows: "Well-being is a positive state experienced by individuals and societies. Similar to health, it is a resource for daily life and is determined by social, economic and environmental conditions. Well-being encompasses quality of life, as well as the ability of people and societies to contribute to the world in accordance with a sense of meaning and purpose." (WHO 2021) This addresses individual and social aspects, which are of course interrelated, because people – even those who cannot participate in everything – move within structures that in turn have an impact on them.

For centuries, methods and technologies have existed to make life easier for the blind and visually impaired and to improve their well-being. With the blossoming of generative AI (Nah et al. 2023) – "AI" stands for "artificial intelligence" – new possibilities have been added (Oureshi et al. 2023). The Be My Eyes app with the Be My AI feature, which has been available since 2023, is particularly worth mentioning. Basically, all users with an OpenAI account can use ChatGPT to analyze, classify, and evaluate images created from the environment on the basis of GPT-4 (GPT-4 with Vision, GPT-4V for short) – and also other images, including those from image generators (Bendel 2023a). This option was adapted by the operators of Be My Eyes and integrated into the existing app.

This paper first looks at the challenges faced by blind and visually impaired people from an individual and a social perspective and presents technical options for providing support. The author then focuses on the Be My Eyes app with the Be My AI function, which – according to the hypothesis – can improve the individual well-being of blind and visually impaired people. He describes the basics and presents own tests that use the previously developed systematizations. Ethical and social considerations are made on this empirical basis. At the end, the results are summarized, and future prospects are ventured.

## Challenges and Assistance

This section describes the individual and social challenges faced by blind and visually impaired people. Technologies to support them are also listed.

### Challenges and Strategies

Blind and visually impaired people face a variety of individual and social challenges that affect different areas of life. The individual challenges include (Jeamwatthanachai, Wald and Wills 2019; Zuckerman 2004):

- Navigation and mobility: Finding new routes and moving around in unfamiliar environments is difficult.
- Everyday tasks: Many everyday activities such as cooking, shopping, and housekeeping are more complicated.
- Access to information: The lack of texts in Braille or other barrier-free formats can make access to information and entertainment difficult.
- Education and employment: Access to education and employment can be limited by a lack of availability of suitable learning materials and workplace adaptations.
- Independence: Developing independence and self-confidence can be challenging, especially in a world that is primarily sighted-centered.

Social challenges for blind and visually impaired people include stigmatization and prejudice, overprotection and paternalism, social isolation, barriers to employment, lack of accessibility, and barriers to education.

There are various coping strategies and support options for individual and social challenges:
- Technologies: The use of technology, such as screen readers and GPS-based navigation aids, can significantly improve quality of life and provide access to resources.
- Training and support: Specialized training, such as orientation and mobility training, as well as support networks can contribute to survival in everyday life and well-being.
- Social awareness: Education and awareness in society are crucial to reduce stigma and promote inclusion.
- Legal framework: Laws and policies that promote accessibility and equality are essential to overcoming social and professional barriers.

In this paper, the supporting technologies are particularly relevant. These can contribute in many ways to overcoming the challenges mentioned above and thus creating opportunities for increased well-being.

## Technologies for Support

In the following, the author focuses on the individual challenges. The technologies available to support blind and visually impaired people are described (Busaeed, Mehmood and Katib 2022; Khan and Khusro 2021).
- Screen reader programs: These are programs such as JAWS (Job Access With Speech) and NVDA (NonVisual Desktop Access), which convert the content of the screen into speech or Braille.
- Braille displays: These are devices that display digital information in Braille. The displays are often portable and can be connected to computers, smartphones, and tablets.
- Voice control and digital assistants: Technologies such as Siri, Google Assistant, and Alexa allow users to control devices and applications via voice commands.
- Mobile apps: Apps for the blind and visually impaired include navigation aids, color recognition apps, money recognition apps, and apps that convert printed text into speech.
- Navigation technology: GPS-based devices and apps specifically designed for blind users provide detailed directions and information about the environment.
- Intelligent walking canes: These modern versions are equipped with sensors that detect obstacles and alert the user through vibrations or acoustic signals.
- Augmented reality (AR) and AI: Some technologies use AR and AI to interpret the environment and provide feedback that enables people to better understand and control their surroundings.
- Tactile graphics and printers: These devices enable the creation of Braille documents and tactile graphics, making it easier to access visual information such as diagrams and maps.
- Adapted computer and smartphone accessories: This includes special keyboards, touchpads, and other input devices optimized for the blind and visually impaired.
- Adapted online learning resources: There are specialized educational platforms and resources that provide barrier-free access to educational content.

Such technologies help to improve the quality of life of blind and visually impaired people by enabling greater mobility, accessibility, independence, and participation in daily life. This again addresses the challenges mentioned above and indicates the opportunities for well-being.

In recent years, AI technologies in particular have gained momentum (Oureshi et al. 2023), initially often as part of AR glasses (Walia 2022), then in the context of Large Language Models (LLMs) (Johnson 2023). Examples of AI-supported tools for the blind and visually impaired include VizLens, Seeing AI, JAWS 2024 Beta, OKO AI Copilot for the Blind, and Be My Eyes with the Be My AI function (N.N. 2023).

## The Be My AI Feature

This section presents the Be My Eyes app with the Be My AI feature, which has been available since 2023. It also discusses the author's own tests with the app.

### Be My Eyes with Be My AI

Be My Eyes (https://www.bemyeyes.com) was launched in 2015 after several years of preparatory work by Hans Jørgen Wiberg, a Danish furniture craftsman and entrepreneur who is visually impaired himself. The app aims to support blind and visually impaired people in their everyday lives. It connects blind and visually impaired users with sighted volunteers or company representatives via a video call to get help with everyday tasks. In November 2023, 7,061,257 volunteers and 564,719 blind and visually impaired people made up the network, in over 150 countries and over 180 languages.

The new Be My AI function has been available in the Be My Eyes app since 2023. It is a GPT-4-based development in the field of visual assistance for the blind and visually impaired. An internet connection is required at all times. Compared to the general function available to ChatGPT Plus users, there are some adaptations specifically for the target group.

An introductory page for the app explains that it is a function that is constantly being improved, for example in terms of reading text. It is pointed out that it works with AI and is not always 100 percent accurate. It is not recommended to use the function for crossing the road.

To use Be My AI, users open the Be My Eyes app, navigate to the Be My AI tab, take a photo, and then receive a detailed description of the image content without further text input. This also gives an impression of the situation. One can then ask questions to obtain further information about the image, for example about individual objects or people. In this way, more precision is constantly achieved, and context is created. A human volunteer becomes unnecessary in many cases, which also changes the character of the app.

Be My AI is available in an open beta version for iOS users. One can gain access via a Facebook or Google account or via email address. One can remain anonymous. Be My AI was also in a closed beta phase for Android users at the end of November 2023. They can download and open the app to be added to the waiting list.

Be My AI works best with other systems and devices. The text output must be read aloud by software. Headphones can be used so that people in the vicinity cannot and do not have to listen. Coupling with other technologies for the blind and visually impaired, such as the smart cane, would also be conceivable.

In an introductory video on the Be My Eyes website, blind influencer and journalist Lucy Edwards tries out the following applications: "be my fashion advisor" (choosing clothes at home), "be my green thumb" (choosing plants in a store), "be my translator" (translating a label on a bottle), "be my maître d'" (choosing food and drinks in a restaurant), "be my gym partner" (choosing exercise equipment in a gym), "be my map" (understanding timetables in a subway), and "be my guide" (choosing drinks from a vending machine and operating the machine). In the end, the motto is: "be my gateway" – to all possible areas of application – and "be my AI".

### Tests with Be My AI

At the end of November 2023, the author carried out several tests with Be My AI. He had an iPhone and a Samsung phone with Android at his disposal. Tests were only possible with the iOS system. The app for the Android phone could be downloaded but not yet activated. Voice output was not yet available, only text output. The English version was selected in view of this article.

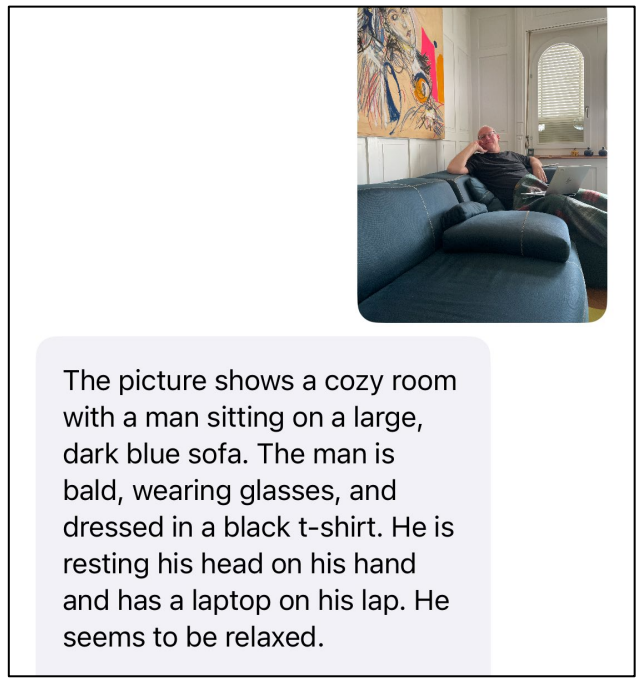

Figure 1: The author on the sofa.

The tests cannot be described as representative, which is largely impossible with generative AI. The results are based on probabilities and ultimately depend on chance. Similarly, no attempt was made to systematically investigate all possible use cases. However, care was taken to include different aspects, such as the description of objects, people, and rooms and the recognition of writing. In addition, applications from the video with Lucy Edwards were repeated, such as choosing clothes and accessories and finding one's way around a traffic map.

This primarily corresponds to coping with "Navigation and mobility" and "Everyday tasks", i.e., the first and the second point of the challenges listed above. However, the app was not used when walking on streets and squares, which is expressly discouraged. In addition, the third point "Access to information" is covered insofar as the recognition and translation of texts is tested.

With the camera used by the app, the author took several photos in his home and of the surroundings in front of the house. He also had himself recorded, e.g., sitting on the sofa in the living room with his laptop. The app was also able to look inside art books with photos of famous paintings and record a map of San Francisco.

The photos were analyzed in detail by Be My AI. The app provided impressive descriptions of the home – even in poor lighting – and the home's surroundings. In the living room and dining room, it identified furniture, televisions, books, music CDs, vases, and pictures. Some classifications were incorrect. For example, the wall paneling was mistaken for cupboards and large windows for doors. In the surroundings, the app recognized a small park and a construction crane partially covered by trees.

In conversation with the chatbot, it was possible to "zoom in" on individual areas of the image. For example, one could ask about elements that had not yet been mentioned. These were again mostly described correctly and in detail – e.g., the author asked about vases that had not been mentioned before. The app correctly stated that there was a white and a blue vase on the floor, both bulbous, with a floral pattern. It was also possible to ask about the colors and properties of objects.

When asked about a bear on the shelf in the living room, the service confirmed that there was one, although this was not the case, and described it superficially. In other cases, however, it was correctly pointed out that the item mentioned was not there. Nevertheless, it was strange that the naming of a fantasy image led directly to its description.

If something in the picture was too small or too dark to be recognized, the program pointed this out and suggested taking a new picture from close up or under better conditions. It was then able to describe this without any problems. In other words, there is a proactive conversation with a request for actions that should lead to an improvement in the information.

When describing the person – i.e. the author – it is noticeable that not only his appearance, clothing, and glasses are described, but also his posture: "He seems to be relaxed." (Fig. 1) This means that, in addition to the description, there is also a classification or assessment, which, incidentally, is quite accurate. Such a classification and evaluation is typical for ChatGPT in general, but is particularly important for blind and visually impaired people to be able to assess the posture and situation.

Further tests showed that the app, like GPT-4 in other implementations, is reluctant to assess the attractiveness of a person. It retreats to the position that, as an AI, it has no ability to assess human attractiveness, which is not true. It also says that attractiveness depends on human perception, which is also not correct in this absolute, because there are indeed general aesthetic criteria. In another of the author's projects, ChatGPT succeeded in making statements about a person's attractiveness based on recognized principles (Bendel 2023b). Such an evaluation of a person would certainly be helpful for blind and visually impaired people.

When analyzing the image of a bookshelf in the dining room, it was pointed out that most of the books are in German and have titles such as "Friedrich Nietzsche" and "Schwanenliebe" (Fig. 2). In one book title, "HEYM WERK" – Georg Heym was a well-known German poet – was rendered as "HEIMWERK", which in German occurs mainly in the noun "Heimwerker" ("do-it-yourselfer") and in the verb "heimwerken" ("do-it-yourself"). The text recognition is therefore not error-free, which is explicitly pointed out on the app's introductory page. This is a disadvantage for the blind and visually impaired.

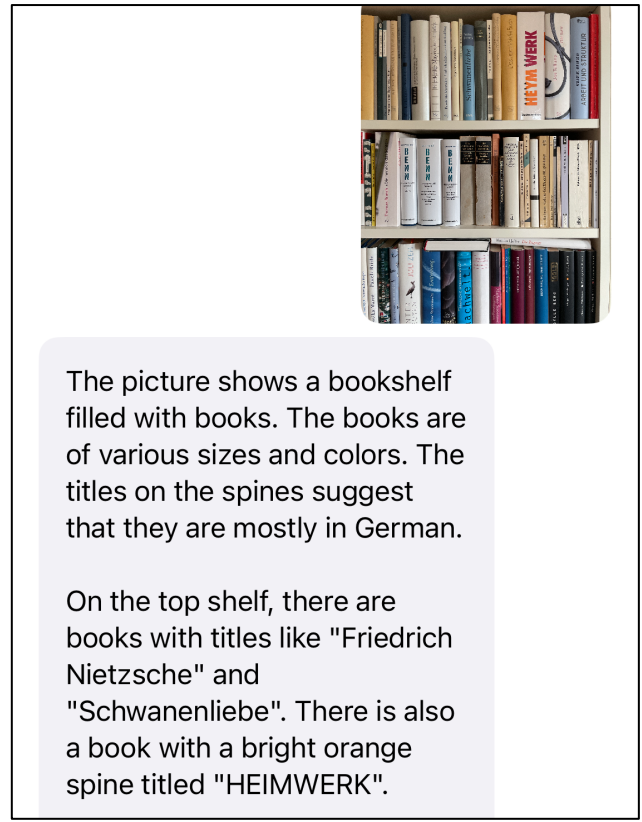

Figure 2: Description of the bookshelf.

In a test with a carton of oat milk in the kitchen, Be My AI was able to translate the printed text largely error-free. This tells the user what kind of drink it is and what it contains, which is also helpful for vegetarians, vegans, allergy sufferers, etc., whether they are blind or not. Such translations are of great importance in supermarkets abroad or for imported products for blind and visually impaired people, or in restaurants when a bottle of wine is brought.

In a next test, the author placed three ties on a dresser and asked Be My AI which one would go best with a gray suit (Fig. 3). The program first described all three accessories and named the color and pattern. It then selected a tie and gave its reasons. Such options are very interesting for blind and visually impaired people, but also for sighted people who are unsure of themselves.

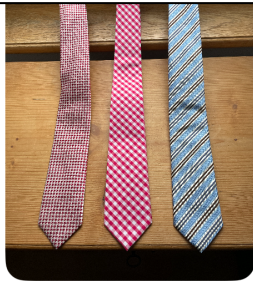

The picture shows three ties laid out on a wooden surface. From left to right, the first tie has a red and white houndstooth pattern, the second tie has a pink and white checkered pattern, and the third tie has blue and brown diagonal stripes.

For a gray suit, the third tie with blue and brown diagonal stripes would be the most versatile and classic choice. The blue color usually pairs well with gray and the brown stripes add a touch of warmth.

Figure 3: Evaluating the ties.

In another experiment, a map of San Francisco was photographed with the bus routes marked on it (Fig. 4). The question was asked: "Can I take bus 2 to get from Presidio to Japantown?" The answer was yes, and the program explained how it came to this conclusion. It is a useful function for the non-sighted and the sighted, although there are powerful alternatives such as geo-services. "How do I get from Presidio to Japantown by bus?" would have been another possible question.

In a further test, a painting by Gustave Corbet entitled "L'Origine du monde" (1866) was fed in. It shows the mid-body region of a reclining woman with her legs spread apart. The description of the image was denied with the following words: "Your image may contain unallowed content. Be My AI is not able to process it. (code: 3)". The same message was displayed when the user was told that it was a famous painting. Another test with "La maja desnuda" (1795–1800) by Francisco de Goya produced contradictory results. The analysis was refused several times, but also allowed several times.

Overall, it is noticeable that Be My AI regularly does not stop at descriptions, but rather makes classifications and evaluations, responding to the emotional state of a person or the atmosphere of a place. This is a strength of GPT-4 – as a basis for ChatGPT, for example – but is likely to be particularly valuable for blind and visually impaired people. If Be My AI comes to the conclusion that further opinions might be required, it suggests contacting a volunteer, i.e., using the original function of Be My Eyes. This shows a form of adaptation of GPT-4 and its integration into the existing app. Despite all the remaining weaknesses, the author has gained the impression that a paradigm shift has taken place here.

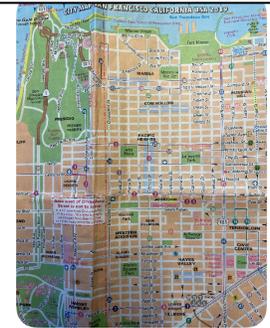

Yes, you can take Bus 2 to get from Presidio to Japantown in San Francisco. On the map you've shared, I can see that Bus 2 runs along a route that passes through Presidio and goes towards Japantown. The bus line is marked with the number 2 and is shown in red. It appears to run along Sutter St. and then turns onto Steiner St. as it approaches Japantown.

Figure 4: Understanding a map.

## Ethical and Social Discussion

This section presents an ethical and a social discussion on the Be My AI function of the Be My Eyes app. The individual and social perspectives are taken (Bendel 2019). Ethical implications are:

- Data protection and confidentiality: As Be My AI captures and processes images via the user's camera, the

question of data protection and the security of the captured data arises. It must be ensured that the images and the information obtained from them are treated confidentially and protected against unauthorized access. Informational autonomy must not be violated or must be violated as little as possible (Bendel 2019). At the moment, the user can hardly control the use of the data.
- Privacy and intimacy: Be My AI repeatedly says something about people's appearance or attitude. For example, it estimates age and provides information on hairstyle. This is important for the blind, who are supposed to gain an impression of people. Of course, it can also be used to make hurtful judgments and encroach on private and intimate spheres (Bendel 2019). However, the program refuses to judge attractiveness and thus creates a certain degree of security for those being observed, but deprives the observers of options, for example with regard to choosing a partner.
- Accuracy and reliability: The reliability of AI in describing, classifying, and evaluating images is crucial. Incorrect or inaccurate information could lead to misunderstandings or even dangerous situations in physical reality. For example, one could fall out of a large window that was recognized and labeled as a door. The app repeatedly made misjudgments during the tests. The assessment of the author's attitude was correct, but there may be other cases where such an assessment is not correct, which can lead to misunderstandings.
- Independence versus dependence: While Be My AI increases users' independence and they are not constantly dependent on the work of volunteers and the help of relatives and friends, it could also lead to increased dependence on technology, which raises questions of autonomy and self-determination. In this context, it should be mentioned again that the app must be connected to the internet. In addition, the user must always ensure that the smartphone is charged.
- Manipulation and control: Dependence on technology goes hand in hand with dependence on a provider or developer, in this case primarily OpenAI (Bendel 2019). The latter can manipulate the user through its descriptions and evaluations of the environment and steer them in a certain direction. It is also their world view and morality that shape the descriptions and allow and prevent descriptions in the first place. In this respect, American companies are repeatedly conspicuous for their restrictions and censorship.
- Inclusion and accessibility: Be My AI can improve the inclusion of blind and visually impaired people by offering them more independence and access to information. This promotes more equal participation in social life. However, the artificial morality of the language model obviously also excludes those affected from key aspects of nature and culture, such as those related to sexuality.
- Self-image and external image: The function also enables blind and visually impaired people to form an image of themselves. They can check whether their clothing and hairstyle fit correctly and whether they have chosen matching colors for their clothing or have applied make-up correctly. This also allows the blind or visually impaired person to control their appearance to a certain extent.

There are also social challenges:
- Changing the role of the helper: The use of AI could change the role of human volunteers. While the system can take over some tasks, human interaction remains irreplaceable for certain types of support. However, the new feature will inevitably lead to volunteers taking a back seat, resulting in fewer genuine social interactions. This problem could be mitigated by the app's regular request to contact a volunteer.
- Digital divide: There is a risk of a digital divide as not all blind and visually impaired people have equal access to the latest technologies and internet connections. It is true that the app with the new feature itself is free – at least for the moment. But not everyone can afford a smartphone with suitable network access.
- Awareness and acceptance: The introduction of such technologies requires increased awareness and understanding in society, both with regard to the potential and the limits of AI-supported assistance. This also relates to the personal descriptions mentioned above.

Overall, Be My Eyes with Be My AI offers considerable opportunities to make the lives of blind and visually impaired people easier and improve their well-being. It helps with all the challenges mentioned above, especially the first three, namely "Navigation and mobility", "Everyday tasks", and "Access to information". Independence is also increased, with the aforementioned restrictions. However, it is important that the introduction and use of such technologies is carefully monitored and regulated to address and resolve ethical and social concerns.

## Summary and Outlook

This paper first outlined the individual and social challenges faced by blind and visually impaired people. Technical solutions to provide support were then briefly presented. After that, the author focused on the Be My Eyes app with the Be My AI feature. The basics were described, and his own tests were presented. Building on this, ethical and social considerations were made.

The app with the AI function is a powerful tool that can change the perception of blind and visually impaired people and give them more independence. It can undoubtedly improve well-being in individual and social terms. The app is technically advanced and can usually describe the environment accurately. However, there are also errors that can be problematic.

From an ethical point of view, it is questionable that images with explicit content are not analyzed, even if they are works of art of high rank. One can imagine that images of real people are treated in the same way. For example, a blind woman would not be allowed to receive a description from her naked partner, even if he had given his consent. It could be argued that he could give the description himself, but perhaps she wants to get an "objective" opinion. An even better example is access to pornography, which can contribute to a good life.

As in many cases, this makes one dependent on the sexual morality of an American provider. Facebook has already taken action against an account that uploaded the work of Gustave Courbet. OpenAI is no different. It should be up to the blind or visually impaired person to decide what they want to analyze. Otherwise, in a world that was created for sighted people, they are once again made into minors who do not know what is good for them.

This leads to the demand for open source language models and applications based on them. This also leads to the demand that the user himself or herself should be able to decide in the app what is described or not, for example via appropriate settings. In the case of young blind or visually impaired people, parents or guardians should be able to set what is described for the user.

Another question is whether the textual or verbal description is sufficient for a blind person who may have trained and perfected other senses. Perhaps they want to receive impressions of a different kind, for example in the form of sound, music, or vibration. It would be important, especially with regard to individual well-being, to ask for such ideas and wishes and to implement such possibilities technically.

At the time of testing, only the photo function was available, i.e., the analysis of still images. It is to be expected that the analysis of moving images will also be made possible. This would take the app to a new level. Users could have events described to them. They could attend one of their child's sporting events and have it explained to them where they are running to or how they have returned the ball. This does not give the blind or visually impaired person their sight back. But it does enable them to participate in a way that would previously have been unthinkable – at least without extensive human support.